\documentclass[prl,aps,10pt, twocolumn, superscriptaddress, floatfix, nofootinbib, amsmath, amssymb, preprintnumbers]{revtex4-1}

\usepackage{dcolumn}
\usepackage{verbatim}

\usepackage{breakurl}

\usepackage{lmodern}

\pdfoutput=1
\usepackage{graphicx}
\usepackage{bm}
\usepackage{amsmath}
\usepackage{amsfonts}
\usepackage{amssymb}
\usepackage{multirow}
\usepackage[colorlinks,linktocpage,linkcolor=cyan,citecolor=cyan]{hyperref}
\usepackage{adjustbox}
\usepackage{array}
\usepackage[capitalise]{cleveref}
\usepackage{acro}
\usepackage[utf8]{inputenc}
\usepackage{CJKutf8}
\usepackage{amssymb}
\usepackage{subfigure}

\usepackage{tikz}
\usepackage{pgfplots}
\pgfplotsset{compat=newest,every axis plot/.append style={line width=1pt}}

\crefname{figure}{Fig.}{Figs.}
\Crefname{figure}{Fig.}{Figs.}
\def\({\left(}
\def\){\right)}
\def\[{\left[}
\def\]{\right]}

\newcommand{\be}{{\begin{eqnarray}}}
\newcommand{\ee}{{\end{eqnarray}}}

\newcommand{\cA}{\mathcal{A}}
\newcommand{\cN}{\mathcal{N}}

\newcommand{\cP}{\mathcal{P}}

\newcommand{\cS}{\mathcal{S}}
\newcommand{\cO}{\mathcal{O}}

\newcommand{\bk}{\mathbf{k}}

\newcommand{\bx}{\mathbf{x}}

\newcommand{\ud}{\mathrm{d}}

\newcommand{\uGW}{\mathrm{gw}}

\newcommand{\Beq}{\begin{align}}
\newcommand{\Eeq}{\end{align}}

\DeclareAcronym{BH}{
  short = BH ,
  long = black hole ,
  short-plural = s ,
}
\DeclareAcronym{SNR}{
  short = SNR ,
  long = signal-to-noise ratio ,
  short-plural = s ,
}
\DeclareAcronym{IMRPPv2}{
  short = ,
  long = {\normalsize IMRP}{\footnotesize HENOM}{\normalsize P}v2 ,
  short-plural = ,
}
\DeclareAcronym{SFR}{
  short = SFR ,
  long = star formation rate ,
  short-plural =  ,
}
\DeclareAcronym{IMR}{
  short = IMR ,
  long = inspiral-merger-ringdown ,
  short-plural =  ,
}
\DeclareAcronym{ABH}{
	short = ABH ,
	long  = astrophysical black hole,
  short-plural = s ,
}
\DeclareAcronym{GW}{
  short = GW ,
  long = gravitational wave ,
  short-plural = s ,
}
\DeclareAcronym{GWB}{
  short = GWB ,
  long = gravitational wave background ,
  short-plural = s ,
}
\DeclareAcronym{SGWB}{
  short = SGWB ,
  long = stochastic gravitational-wave background ,
  short-plural = s ,
}
\DeclareAcronym{CBC}{
  short = CBC ,
  long = compact binary coalescence ,
  short-plural = s ,
}
\DeclareAcronym{BBH}{
  short = BBH ,
  long = binary black hole ,
  short-plural = s ,
}
\DeclareAcronym{PBH}{
  short = PBH ,
  long = primordial black hole ,
  short-plural = s ,
}
\DeclareAcronym{LIGO}{
  short =LIGO ,
  long = Laser Interferometer Gravitational-Wave Observatory ,
  short-plural = ,
}
\DeclareAcronym{LVK}{
  short = LVK ,
  long = {LIGO, Virgo and KAGRA} ,
  short-plural = ,
}
\DeclareAcronym{ET}{
	short = ET ,
	long  = Einstein Telescope,
  short-plural =  ,
}
\DeclareAcronym{CE}{
	short = CE ,
	long  = Cosmic Explorer,
  short-plural =  ,
}
\DeclareAcronym{LISA}{
	short = LISA ,
	long  = Laser Interferometer Space Antenna,
  short-plural =  ,
}
\DeclareAcronym{BBO}{
	short = BBO ,
	long  = Big Bang Observer,
  short-plural =  ,
}
\DeclareAcronym{DECIGO}{
	short = DECIGO ,
	long  = Deci-hertz Interferometer Gravitational wave Observatory,
  short-plural =  ,
}
\DeclareAcronym{PTA}{
  short = PTA ,
  long = pulsar timing array ,
  short-plural = s ,
}
\DeclareAcronym{FRW}{
  short = FRW ,
  long = Friedman-Robertson-Walker ,
  short-plural =  ,
}
\DeclareAcronym{CMB}{
  short = CMB ,
  long = cosmic microwave background ,
  short-plural =  ,
}
\DeclareAcronym{SSR}{
  short = SSR ,
  long = sound speed resonance ,
  short-plural =  ,
}
\DeclareAcronym{SIGW}{
  short = SIGW ,
  long = scalar-induced gravitational wave ,
  short-plural = s ,
}
\DeclareAcronym{SKA}{
  short = SKA ,
  long =  Square Kilometer Array ,
  short-plural =  ,
}
\DeclareAcronym{NANOGrav}{
  short = NANOGrav ,
  long =  North American Nanohertz Observatory for Gravitational Waves ,
  short-plural =  ,
}
\DeclareAcronym{PNG}{
  short = PNG ,
  long =  primordial non-Gaussianity
 ,
  short-plural =  ,
}
\DeclareAcronym{AD}{
  short = AD ,
  long =  Affleck-Dine
 ,
  short-plural =  ,
}
\DeclareAcronym{SM}{
  short = SM ,
  long =  Standard Model
 ,
  short-plural =  ,
}
\DeclareAcronym{SUSY}{
  short = SUSY ,
  long =  supersymmetry
 ,
  short-plural =  ,
}
\DeclareAcronym{eRD}{
  short = eRD ,
  long =  early radiation-dominated
 ,
  short-plural =  ,
}
\DeclareAcronym{eMD}{
  short = eMD ,
  long =  early matter-dominated
 ,
  short-plural =  ,
}
\DeclareAcronym{RD}{
  short = RD ,
  long =  radiation-dominated
 ,
  short-plural =  ,
}
\DeclareAcronym{VEV}{
  short = VEV ,
  long =  vacuum expectation value
 ,
  short-plural =  s,
}

\begin{document}

\title{Large Anisotropies in the Gravitational Wave Background from Baryogenesis}

\author{Yan-Heng Yu}
\email{yhyu@ihep.ac.cn}
\affiliation{Theoretical Physics Division, Institute of High Energy Physics, Chinese Academy of Sciences, Beijing 100049, China}
\affiliation{School of Physical Sciences, University of Chinese Academy of Sciences, Beijing 100049, China}

\author{Sai Wang}
\email{Corresponding author: wangsai@ihep.ac.cn}
\affiliation{School of Physics, Hangzhou Normal University, Hangzhou 311121, Zhejiang, China}
\affiliation{Theoretical Physics Division, Institute of High Energy Physics, Chinese Academy of Sciences, Beijing 100049, China}

\begin{abstract} 
Affleck-Dine (AD) baryogenesis can produce the baryon asymmetry of the Universe through the $CP$-violating dynamics of AD field. The field generally fragments into Q-balls, whose rapid decay induces enhanced gravitational waves. In this Letter, we investigate the anisotropies in this gravitational wave background as a new essential observable for AD baryogenesis. The evolution of AD field causes non-Gaussian baryonic isocurvature perturbations, and the non-Gaussianity modulates the spatial distribution of Q-balls on large scales, resulting in large-scale anisotropies in the Q-ball-induced gravitational wave background. We present that the anisotropies can be significantly large with a reduced angular power spectrum $\sim 10^{-2}$, and can be detected by future experiments like LISA. Moreover, these anisotropies universally reveal the $CP$-violating dynamics of AD field, opening a novel road to explore the longstanding baryon asymmetry puzzle.
\end{abstract}

\maketitle

\acresetall

\emph{Introduction.}---
The origin of the baryon asymmetry of our Universe remains a fundamental challenge in particle physics and cosmology, unaddressed in the framework of the \ac{SM} \cite{Sakharov:1967dj,Dine:2003ax}.
\ac{AD} baryogenesis is a well-established mechanism to produce the observed baryon asymmetry, driven by the evolution of an \ac{AD} field, which can be a flat direction with a non-zero baryon number in extensions of the \ac{SM} \cite{Affleck:1984fy,Dine:1995kz,Enqvist:2003gh,Allahverdi:2012ju}.
The \ac{AD} field forms scalar condensates during inflation, and the post-inflationary dynamics of the condensates efficiently generate baryon density through $CP$-violating interactions. 
Since this mechanism is expected to happen at high energy scales inaccessible to particle colliders, searching for corresponding cosmological observables becomes a basic task to test \ac{AD} baryogenesis and to explore the dynamics of \ac{AD} field.

A promising observational window for \ac{AD} baryogenesis is Q-balls in the early Universe.
Generic models predict that \ac{AD} condensates fragment into non-topological solitons termed Q-balls \cite{Coleman:1985ki,Lee:1991ax}, whose stability is protected by the conserved charge (i.e., baryon number).
Q-balls have garnered increasing attentions for their profound cosmological implications, e.g., they can dominate and reheat the early Universe \cite{Enqvist:2002rj,Enqvist:2002si}, constitute all the dark matter \cite{Kusenko:1997si,Kusenko:1997vp,Kusenko:2001vu,Kasuya:2001hg}, and provide seeds for \acp{PBH} \cite{Cotner:2016cvr,Cotner:2017tir,Cotner:2019ykd,Flores:2021jas}.
The properties of Q-balls, which are determined by details of the evolution of \ac{AD} field, naturally carry rich information about \ac{AD} baryogenesis.

\Acp{GW}, which propagate almost freely since their generation in the early Universe, offer a direct probe to cosmological scenarios involving Q-balls. 
Various processes associated with Q-balls can be sources of \acp{GW}, including their formation \cite{Kusenko:2008zm,Kusenko:2009cv,Chiba:2009zu,Zhou:2015yfa}, decay \cite{White:2021hwi,Kasuya:2022cko,Kawasaki:2023rfx,Flores:2023dgp}, and universal statistical properties \cite{Lozanov:2023aez,Lozanov:2023rcd,Luo:2025lgr}. 
This work concentrates on the \acp{GW} related to Q-ball decay in the context of \ac{AD} baryogenesis, whose frequency band can be covered by a wide range of \ac{GW} detectors.  
Since Q-balls redshift as matter, their energy density can exceed radiation's, leading to an \ac{eMD} era.
When Q-balls decay into \ac{SM} particles, with a typical timescale much shorter than the Hubble time, the Universe transition into a \ac{RD} era.
As the transition is sudden, Q-ball density perturbations reentering the horizon during the \ac{eMD} era have no time to sufficiently dissipate. 
They evolve into relativistic sound waves in the \ac{RD} era and resonantly induce \acp{GW} \cite{Inomata:2019ivs}, acting as detectable signals for \ac{AD} baryogenesis.
Existing studies focus on the spectrum of these \acp{GW} \cite{White:2021hwi,Kasuya:2022cko,Kawasaki:2023rfx,Flores:2023dgp}.
However, the \ac{GW} spectrum solely provides limited information about the dynamics of \ac{AD} field.
Besides, other cosmological scenarios involving \acp{PBH} \cite{Inomata:2020lmk,Papanikolaou:2020qtd,Domenech:2020ssp,Domenech:2021wkk,Papanikolaou:2022chm,He:2024luf,Domenech:2024wao}, oscillons \cite{Lozanov:2022yoy,Sui:2024grm}, curvatons \cite{Kumar:2024hsi}, or axions \cite{Harigaya:2023mhl}, etc., can also produce similar \ac{GW} spectra.

In this Letter, we propose the anisotropies in the \ac{GWB} as a new essential observable for \ac{AD} baryogenesis.
In \ac{AD} baryogenesis, the quantum fluctuations of \ac{AD} field during inflation can cause non-Gaussian baryonic isocurvature perturbations through $CP$-violating interactions \cite{Kawasaki:2008jy}.
The spatial distribution of resulting Q-balls can be affected by this non-Gaussianity, leading to non-linear couplings between the large- and small-scale Q-ball density perturbations.
We will demonstrate that the \acp{GW} induced by small-scale Q-ball density perturbations would experience modulation on superhorizon scales due to the presence of large-scale perturbations, resulting in observable anisotropies in the \ac{GWB}.
These anisotropies are revealed to exhibit close connections to the $CP$-violating dynamics of \ac{AD} field.

Furthermore, we will present that the anisotropies in the \ac{GWB} from \ac{AD} baryogenesis can be remarkably large, with a reduced angular power spectrum $\sim 10^{-2}$.
Considering the anisotropies in the \ac{GWB} originated from most cosmological scenarios are challenging to observe, this highly anisotropic \ac{GWB} is of particular observational interests, potentially detected by future \ac{GW} experiments, e.g., \ac{LISA} \cite{LISA:2017pwj,Baker:2019nia,LISACosmologyWorkingGroup:2022jok}, \ac{ET} \cite{Punturo:2010zz}, \ac{CE} \cite{Reitze:2019iox}, \ac{DECIGO} \cite{Seto:2001qf,Kawamura:2020pcg}, \ac{BBO} \cite{Crowder:2005nr,Smith:2016jqs}, etc.
The large anisotropies in the \ac{GWB}, accompanied with the \ac{GW} spectrum, provide distinctive signals for \ac{AD} baryogenesis and valuable insights into the longstanding question of baryon asymmetry.

\emph{Affleck-Dine baryogenesis.}---
The \ac{AD} field for baryogenesis could naturally be many potential-vanishing directions carrying baryon number in supersymmetric theories, parameterized as $\phi=|\phi|\, e^{i\theta}/\sqrt{2}$.
During inflation, $\phi$ forms a scalar condensate with a large vacuum expectation value with $CP$ violation as an initial condition. 
After inflation, as the direction is lifted by some non-renormalizable term (i.e., $\phi^n$ with $n\geq 4$) in superpotential, the condensate relaxes from its initial position and follows a rotating trajectory in field space.
The baryon asymmetry can be effectively created by the angular motion of $\phi$, with the resulting baryon density $n_b \propto \sin{n\theta_\mathrm{i}}$, where $\theta_\mathrm{i}$ denotes its initial value set in inflation.
Here, the phase component $\theta_\mathrm{i}$, manifesting as the $CP$ phase for baryogenesis, is a crucial parameter in this mechanism.

In \ac{AD} baryogenesis, quantum fluctuations of $\theta_\mathrm{i}$ during inflation lead to baryonic isocurvature perturbations $\cS_b \simeq \delta n_b/n_b$ \cite{Enqvist:1998pf,Enqvist:1998en,Kawasaki:2001in,Kasuya:2008xp,Kawasaki:2008jy}.
Given the Gaussian fluctuations $\delta \theta_\mathrm{i}$, $\cS_b$ can be non-Gaussian due to the non-linear dynamics, i.e., $\cS_b=(n\cot{n\theta_\mathrm{i}})\,\delta\theta_\mathrm{i}-(n^2/2)\,(\delta\theta_\mathrm{i})^2$ \cite{Kawasaki:2008jy}. 
It can also be recast as $\cS_b=\cS_{bg} + F_{\mathrm{nl},b}\, \cS_{bg}^2$, where $\cS_{bg}=(n\cot{n\theta_\mathrm{i}})\, \delta\theta_\mathrm{i}$ denotes the Gaussian component of $\cS$, and $F_{\mathrm{nl},b}=-(1/2)\tan^2{n\theta_\mathrm{i}}$ is the non-linearity parameter.
The non-Gaussianity of $\cS_b$ is directly related the the $CP$ phase of $\phi$, and can be significant if $|\tan{n\theta_\mathrm{i}}\,|\gg1$.

\emph{Q-ball density perturbations.}---
Following baryogenesis, the \ac{AD} field condensates generally form Q-balls, with almost all the baryon number being absorbed into these Q-balls \cite{Kusenko:1997si,Kasuya:2000wx}.
Therefore, the initial $\cS_b$ from \ac{AD} baryogenesis are inherited by the resulting Q-ball density perturbations $\cS\simeq \delta_\mathrm{Q}$, where $\delta_\mathrm{Q}$ is the density contrast of Q-balls.
Considering thin-wall Q-balls with the linear mass-charge relation \cite{Coleman:1985ki}, the total energy of Q-balls within a Hubble patch is approximately proportional to their baryon number.
In this case, $\cS$ at a coarse-grained position follow the same statistics as the initial $\cS_b$, i.e.,
\begin{equation}\label{eq:Fnl}
    \cS\simeq\cS_g + F_\mathrm{nl}\, \cS_g^2\ ,
\end{equation}
with $\cS_g\simeq \cS_{bg}$ and $F_\mathrm{nl}\simeq F_{\mathrm{nl},b}$.
The non-linear term in Eq.~(\ref{eq:Fnl}) could introduce the coupling between the large- and small-scale modes in Fourier space through the term $F_\mathrm{nl} \,\cS_{g\mathrm{L}}\cS_{g\mathrm{S}}$.
Here, the subscript ``$\mathrm{L}$'' (or ``$\mathrm{S}$'') represents large (or small) scales, and $F_\mathrm{nl}$ reflects the coupling strength.
It indicates that due to the non-Gaussianity, the spatial distribution of Q-balls at small scales can be modulated across large scales.
Moreover, the statistics of $\cS$ also contains clues of $CP$-violating dynamics of $\phi$ during \ac{AD} baryogenesis. 

The dimensionless power spectrum of $\cS_g$ is defined as $\langle \cS_{g,\bk} \cS_{g,\bk'} \rangle =\delta(\bk+\bk')(2\pi^2/k^3) \cP_{\cS_g}(k)$, where $\langle...\rangle$ represents ensemble average, $\bk$ is the Fourier mode of perturbations and $k=|\bk|$.
At Q-ball formation, $\cP_{\cS_g}$ can be modeled as 
\begin{equation}\label{eq:Sg}
  \cP_{\cS_g}(k)
  = \cA_\mathrm{L}
  +\cA_\mathrm{S}\,\left(\frac{k}{k_\mathrm{max}}\right)^{n_\mathrm{S}}\, \Theta(k_\mathrm{max}-k)\ .
\end{equation} 
On large scales, $\cP_{\cS_g}$ has a nearly scale-invariant amplitude $\cA_\mathrm{L}\simeq  (n\cot{n\theta_\mathrm{i}})^2 \, \cP_{\delta \theta_\mathrm{i}}$, where $\cP_{\delta \theta_\mathrm{i}} \simeq [H_\mathrm{i}/(2\pi|\phi_\mathrm{i}|)]^2$ is given by fluctuations of the light field $\theta$ during inflation, with $|\phi_\mathrm{i}|$ and the Hubble scale $H_\mathrm{i}$ being valued at the horizon exit.
As isocurvature modes, the large-scale amplitude is constrained by the \ac{CMB} observations as $\cA_\mathrm{L} \lesssim 10^{-11}$ \cite{Planck:2018jri}.
On much smaller scales, the statistical characteristics of Q-balls give rise to a blue-tiled spectrum, where $\cA_\mathrm{S}$ is the spectral amplitude at a cutoff scale $k_\mathrm{max}$, and $n_\mathrm{S}$ is the small-scale spectral index. 
For example, the statistically independent Q-ball formation at small scales gives $n_\mathrm{S}=3$, while additional gravitational clustering effects can modify the index to $2\lesssim n_\mathrm{S} \lesssim 3$ \cite{Lozanov:2023aez,Lozanov:2023knf}.

After formation, Q-balls can dominate the early Universe until they decay. 
We assume that the \ac{eMD} era begins at $\eta_\mathrm{m}$ and ends at $\eta_\mathrm{d}$ with $\eta$ the conformal time, and focus on the mode of $\cS$ reentering the horizon during this era, i.e., $k_\mathrm{d}\lesssim k\lesssim k_\mathrm{m}$, where $k_\mathrm{d}\eta_\mathrm{d}=k_\mathrm{m}\eta_\mathrm{m}=1$.
These small-scale $\cS$ evolve into curvature perturbations $\zeta\simeq\cS/3$ and keep constant in the \ac{eMD} era. 
Once Q-balls decay into \ac{SM} particles, their density perturbations transform into relativistic sound waves and act as sources of \acp{GW} in the \ac{RD} era, as will be discussed below. 

\emph{Gravitational waves.}---  
Due to the non-linear effect of gravity, the evolution of $\zeta$ inside the horizon inevitably induces \acp{GW} at the second order \cite{Ananda:2006af,Baumann:2007zm,Mollerach:2003nq,Assadullahi:2009jc,Domenech:2021ztg,Espinosa:2018eve,Kohri:2018awv}.
Taking into account the leading source term, the \ac{GW} strain is given by $h\sim \eta^{2}\zeta'\zeta'$, where a prime denotes the derivation with respect to $\eta$.
For the mode $k\gg k_\mathrm{d}$, $\zeta$ oscillates in the \ac{RD} era at a frequency much larger than the cosmic expansion rate.
Besides, since the \ac{eMD}-to-\ac{RD} transition is rapid, the amplitude of $\zeta$ has no time to decay sufficiently during these oscillations.
These two factors result in amplified \acp{GW} \cite{Inomata:2019ivs}, i.e., $h\sim (k\eta_\mathrm{d})^2 \zeta^2\gg\zeta^2$, as a typical observable for Q-balls in \ac{AD} baryogenesis \cite{White:2021hwi,Kasuya:2022cko,Kawasaki:2023rfx,Flores:2023dgp}.

\begin{figure}[t]
\includegraphics[width=\linewidth]{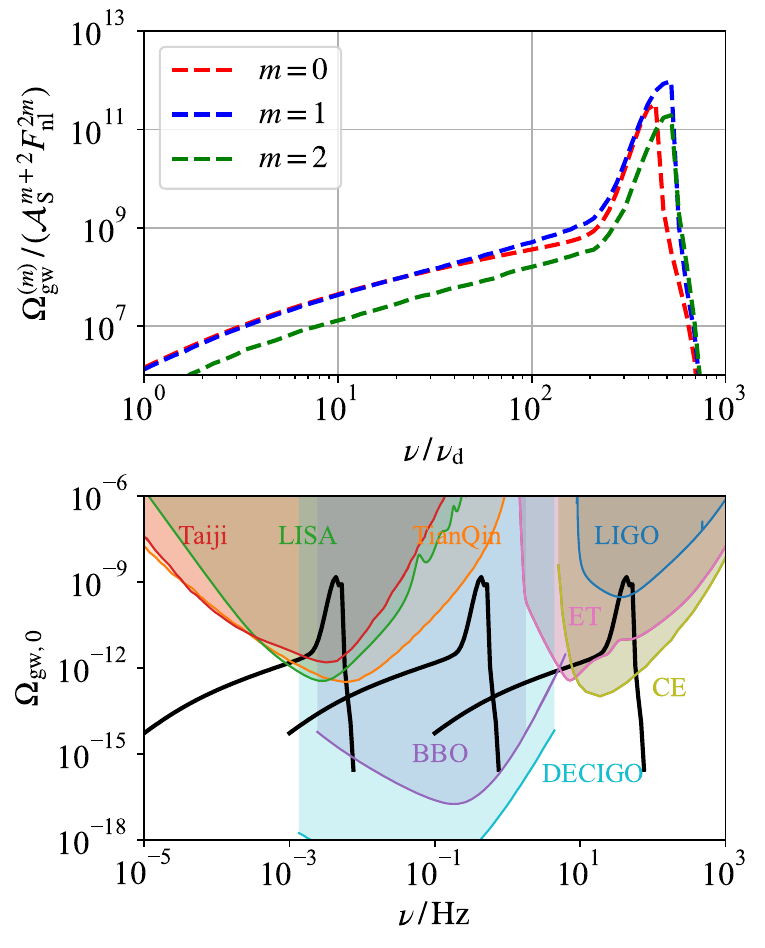}
\caption{
Upper panel: Unscaled components of energy-density fraction spectrum $\Omega_\mathrm{gw}^{(m)}/(\cA_\mathrm{S}^{m+2}F_\mathrm{nl}^{2m})$ versus the \ac{GW} frequency $\nu/\nu_\mathrm{d}$. 
Lower panel: Present-day energy-density fraction spectra $\bar{\Omega}_\mathrm{gw,0}(\nu)$ with $\nu_\mathrm{d}/ \mathrm{Hz}=10^{-5},10^{-3},$ and $10^{-1}$ from left to right.
Other parameters are set as $\cA_\mathrm{S}=5\times 10^{-8}$ and $F_\mathrm{nl}=-2500$.
We compare the spectra to the sensitivities of LISA (green), Taiji (red), TianQin (orange), BBO (purple), DECIGO (cyan), LIGO (blue), ET (pink), and CE (yellow).
In both panels, we assume an instantaneous \ac{eMD}-to-\ac{RD} transition and set $\eta_\mathrm{d}/\eta_\mathrm{m}=400$, $k_\mathrm{max}=k_\mathrm{m}$, and $n_\mathrm{S}=3$.
}\label{fig:1}
\end{figure}

The energy-density fraction spectrum of \acp{GW} at emission is given by $\Omega_\mathrm{gw}(\bx,\nu)=[\ud \ln{\rho_\mathrm{gw}}(\bx)/\ud \ln{\nu}]/\rho_c$, where $\nu=k/(2\pi)$ is the \ac{GW} frequency, $\rho_c$ is the critical energy density of the Universe, and $\rho_\mathrm{gw}(\bx)\sim\langle  \partial h\,\partial h \rangle_\bx$ is the \ac{GW} energy density at a coarse-grained spatial position $\bx$ at emission.
The homogeneous and isotropic component of the \ac{GWB} is described by the background level of $\Omega_{\uGW}(\bx,\nu)$, namely, $\bar{\Omega}_\uGW(\nu)=\langle\Omega_{\uGW}(\bx,\nu)\rangle$. 
In our model, $\bar{\Omega}_\mathrm{gw}$ is proportional to the four-point correlation of the small-scale non-Gaussian $\cS$ (or, equivalently, $\zeta$), i.e., $\cS_{\mathrm{S}}=\cS_{g\mathrm{S}}+F_\mathrm{nl}\,\cS_{g\mathrm{S}}^2$ \cite{Nakama:2016gzw,Garcia-Bellido:2017aan,Cai:2018dig,Unal:2018yaa,Atal:2021jyo,Yuan:2020iwf,Adshead:2021hnm,Ragavendra:2021qdu,Ragavendra:2020sop,Abe:2022xur,Li:2023qua,Li:2023xtl,Yuan:2023ofl,Perna:2024ehx}, and is significantly amplified due to the rapid Q-ball decay, parameterized as
\begin{equation}\label{eq:bogw}
    \bar{\Omega}_\mathrm{gw}(\nu_{\mathrm{S}})=\cN(\nu_{\mathrm{S}}) \,\langle\zeta_{\mathrm{S}}^4\rangle\simeq (1/3)^4\, \cN(\nu_{\mathrm{S}}) \,\langle\cS_{\mathrm{S}}^4\rangle\ .
\end{equation}
Here, $\cN$ is an amplification factor, numerically given by $\cN\sim \mathrm{few}\,\times10^{-7}(\eta_\mathrm{d}/\eta_\mathrm{m})^7$ at $\sim \nu_\mathrm{m}$ \cite{Inomata:2019ivs}.  
In FIG.~\ref{fig:1}, we plot the unscaled results of $\bar{\Omega}_\mathrm{gw}^{(m)}$ $(m=0,1,2)$, where $\bar{\Omega}_\mathrm{gw}^{(m)}$ denotes the $\cO(\cA_\mathrm{S}^{m+2} F^{2m}_\mathrm{nl})$ components of $\bar{\Omega}_\mathrm{gw}$. 
We also plot the present-day energy-density fraction spectrum $ \bar{\Omega}_{\mathrm{gw},0}(\nu_{\mathrm{S}}) \simeq \Omega_{\mathrm{rad}, 0}\, \bar{\Omega}_\mathrm{gw}(\nu_{\mathrm{S}})$, where $\Omega_{\mathrm{rad},0}=4.2\times 10^{-5}/h_0^2$ is the current energy-density fraction of radiation, with $h_0=0.674$ the dimensionless Hubble constant \cite{Planck:2018vyg}. 
The spectrum typically exhibits a sharp peak at $\sim\nu_{\mathrm{m}}$ and can be detected by future \ac{GW} detectors.
Besides, primordial curvature perturbations, with a nearly scale-invarant spectral amplitude $\sim 10^{-9}$ \cite{Planck:2018vyg}, could also be the source of \acp{GW}, but their contribution should be subdominated at $\sim\nu_{\mathrm{m}}$ for $ \cA_\mathrm{S}\gg 10^{-9}$.

\emph{Anisotropies in the gravitational wave background.}---
Let us now focus on the \ac{GW} anisotropies in our model, which arise from the large-scale inhomogeneities in the \ac{GWB}, i.e., $\delta \Omega_{\uGW}(\bx,\nu)=\Omega_\uGW(\bx,\nu)-\bar{\Omega}_\uGW(\nu)$. 
In contrast, we should note that $\delta \Omega_{\uGW}$ on subhorizon scales make no contribution to observed anisotropies. Since the horizon at \ac{GW} emission is extremely small compared to the angular resolution of detectors, the \ac{GW} signal along the line of sight is actually an average over a large number of such horizons.
Therefore, the small-scale $\delta \Omega_{\uGW}$ would vanish after being averaged.

In \ac{AD} baryogenesis, the large-scale $\delta \Omega_{\uGW}$ result from the large-scale Q-ball density perturbations.
As aforementioned, the non-Gaussianity of $\cS$ couples large- and small-scale modes, the latter of which are sources of enhanced \acp{GW}.
Due to the presence of $\cS_{g\mathrm{L}}$, the \ac{GW} energy density  $\sim\langle\cS_\mathrm{S}^4\rangle_\bx$ would be redistributed on large scales  \cite{LISACosmologyWorkingGroup:2022kbp,Malhotra:2022ply,Bartolo:2019zvb,Li:2023qua,Li:2023xtl,Schulze:2023ich,Wang:2023ost,Yu:2023jrs,Ruiz:2024weh,Rey:2024giu}.
The resulting large-scale $\delta\Omega_\uGW$ is proportional to $ \cS_{g\mathrm{L}}$ at the leading order, and can be heuristically expressed as
\begin{equation}\label{eq:dogw}
    \delta\Omega_\mathrm{gw}(\bx,\nu_\mathrm{S}) 
    \simeq 
    (1/3)^4\,\cN(\nu_\mathrm{S})\, F_\mathrm{nl}\,\cS_{g\mathrm{L}}
    \langle \cS_{g\mathrm{S}} \,\cS_\mathrm{S}^3\rangle_\bx \ ,
\end{equation}
where one $\cS_\mathrm{S}$ in the four-point correlation is replaced by the non-linear term $F_\mathrm{nl}\,\cS_{g\mathrm{L}}\,\cS_{g\mathrm{S}}$, compared to Eq.~(\ref{eq:bogw}).
Here, the effects of higher order of $\cS_{g\mathrm{L}}$ is negligible since $\cS_{g\mathrm{L}} \ll  1$.
Though $\cS_{g\mathrm{S}}$ remains uncorrelated for two distant spatial positions, $\cS_{g\mathrm{L}}$ acts as a bridge to correlate large-scale $\delta \Omega_\mathrm{gw}$, leading to $\langle \delta \Omega^2_\mathrm{gw} \rangle\propto\langle \cS^2_{g\mathrm{L}} \rangle \sim  \cA_\mathrm{L}$.

Similar to the \ac{CMB}, the anisotropies in the \ac{GWB} are characterized by the reduced angular power spectrum $\widetilde{C}_\ell$, which is given by the following two-point correlation
\begin{equation}\label{eq:Cldef}
    \ell(\ell+1)\, \widetilde{C}_\ell (\nu)
    =
    \frac{\pi}{2}\, 
    \bigg\langle
    \frac{\delta \Omega_\mathrm{gw}(\bx,\nu)}{\bar{\Omega}_\mathrm{gw}(\nu)}\,
    \frac{\delta \Omega_\mathrm{gw}(\bx',\nu)}{\bar{\Omega}_\mathrm{gw}(\nu)}
    \bigg\rangle \ , 
\end{equation}
where $\ell$ is the multipole corresponding to the angular subtended by two distant positions $\bx$ and $\bx'$. 
Here we omit propagation effects after \ac{GW} emission \cite{Contaldi:2016koz,Bartolo:2019oiq,Bartolo:2019yeu,Schulze:2023ich}.
For large anisotropies of our interest, $\widetilde{C}_\ell$ should be mainly determined by large inhomogeneities at \ac{GW} emission, and propagation effects are negligible in this case.

\begin{figure}[t]
\includegraphics[width=\linewidth]{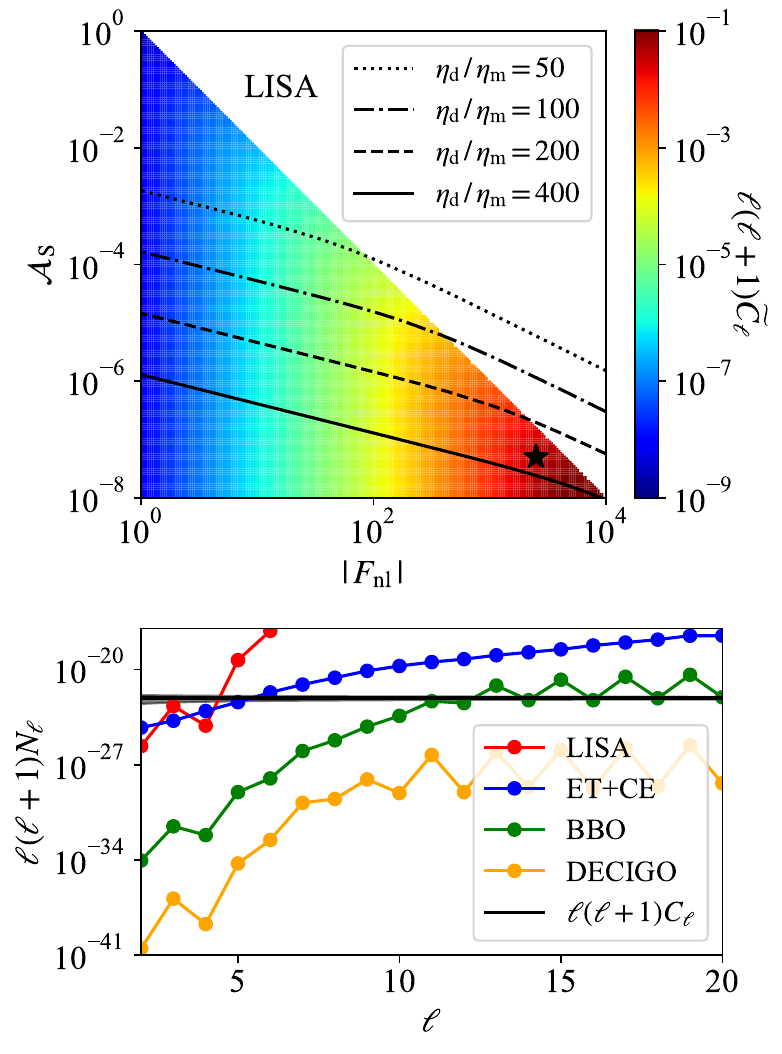}
\caption{
Upper panel: Reduced angular power spectrum $\ell(\ell+1)\, \widetilde{C}_{\ell}(\nu_\mathrm{m}) $ in the parameter space of $\{\cA_\mathrm{S},|F_\mathrm{nl}|\,\}$ with $\cA_\mathrm{L}=10^{-11}$. 
Black lines are given by $\ell(\ell+1)\, C_\ell = 10^{-23}$ for $\eta_\mathrm{d}/\eta_\mathrm{m}=50$ (dotted), $100$ (dashdot), $200$ (dashed), and $400$ (solid).
The regions above these lines are anticipated to be detectable by \ac{LISA}.
Lower panel: Angular power spectrum $\ell(\ell+1)\,{C}_{\ell}(\nu_\mathrm{m}) $ with $\eta_\mathrm{d}/\eta_\mathrm{m}=400$, $\cA_\mathrm{L}=10^{-11}$, $\cA_\mathrm{S}=5\times10^{-8}$, and $F_\mathrm{nl}=-2500$ (the latter two are marked as a star in the upper panel).
Shaded black region represents the uncertainties at 68\% confidence level due to the cosmic variance.
We compare $\ell(\ell+1)\,{C}_{\ell}(\nu_\mathrm{m}) $ to the noise angular power spectra $\ell(\ell+1)\,N_{\ell}$ of \ac{LISA}, \ac{ET}-\ac{CE} network, \ac{BBO}, and \ac{DECIGO}, assuming that $\nu_\mathrm{m}$ aligns with the optimal sensitivity frequency of each detector.
These noise angular power spectra are plotted in Fig.~7 in Ref.~\cite{Braglia:2021fxn}.}\label{fig:C_l}
\end{figure}

Based on Eqs.~(\ref{eq:bogw}-\ref{eq:Cldef}), the reduced angular power spectrum of the \ac{GWB} from \ac{AD} baryogenesis is obtained as
\begin{equation}\label{eq:Cl}
  \ell(\ell+1) \,\widetilde{C}_{\ell}(\nu_\mathrm{S})  
  \simeq
  128\,\pi\cA_\mathrm{L}\, F^2_\mathrm{nl}\, \epsilon^2\ ,
\end{equation}
where $\epsilon=[\bar{\Omega}^{(0)}_\mathrm{gw}+\bar{\Omega}^{(1)}_\mathrm{gw}/2]\,/\bar{\Omega}_\mathrm{gw}$ \cite{Li:2023qua,Li:2023xtl}, measuring the fraction of the \ac{GW} energy density modulated by $\cS_{g\mathrm{L}}$ to the total.
In FIG.~\ref{fig:C_l}, we plot $\ell(\ell+1)\,\widetilde{C}_\ell$ in the parameter space of $\{\cA_\mathrm{S},|F_\mathrm{nl}|\,\}$, and limit our calculation in perturbative regimes (i.e., $\cA_\mathrm{S} F_\mathrm{nl}^2 \ll 1$) to guarantee its validity.
Since $\epsilon\sim 1$ is insensitive to $\cA_\mathrm{S}$ and $F_\mathrm{nl}$, the result of $\widetilde{C}_\ell$ is almost determined by $F_\mathrm{nl}$, i.e., $\widetilde{C}_\ell\propto F_\mathrm{nl}^2$.
Therefore, these anisotropies provide a universal probe to the non-Gaussianity, regardless of details in small-scale $\cP_{\cS_g}$.

To compare with angular sensitivities of \ac{GW} detectors, we define the angular power spectrum of the \ac{GWB} as $C_\ell=[\bar{\Omega}_\mathrm{gw,0}/4\pi ]^2\, \widetilde{C}_{\ell}$.
Specially, we show in the upper panel of FIG.~\ref{fig:C_l} the region in the parameter space $\{\cA_\mathrm{S},|F_\mathrm{nl}|\,\}$ leading to observable anisotropies for \ac{LISA}, say, $\ell(\ell+1)\, C_\ell \gtrsim 10^{-23}$.
For larger $\cN$ (i.e., larger $\eta_\mathrm{d}/\eta_\mathrm{m}$), the region of smaller $\cA_\mathrm{S}$ can be covered by detectors. 

\ac{AD} baryogenesis can generate large anisotropies in the \ac{GWB}, which are also observable for future \ac{GW} detectors.
In this mechanism, Q-balls serve as highly non-Gaussian sources of \acp{GW} and induce the rapid \ac{eMD}-to-\ac{RD} transition, leading to large $|F_\mathrm{nl}|$ and large $\cN$ meanwhile.
As discussed above, these two factors are both important for the observation of large anisotropies.
As a specific realization, we choose the related parameters as $\eta_\mathrm{d}/\eta_\mathrm{m}=400$, $\cA_\mathrm{L}=10^{-11}$, $\cA_\mathrm{S}=5\times10^{-8}$, and $F_\mathrm{nl}=-2500$, which do not violate the perturbative condition or observational constraints from the \ac{CMB}.
Under these choices, the reduced angular power spectrum at $\sim \nu_\mathrm{m}$ is estimated as
\begin{equation}\label{eq:Cl 10-2}
  \ell(\ell+1)\, \widetilde{C}_{\ell}(\nu_\mathrm{m})\,  
  \simeq 0.016 \ .
\end{equation}
In the lower panel of FIG.~\ref{fig:C_l}, we plot the corresponding angular power spectrum $ \ell(\ell+1)\, C_\ell(\nu_\mathrm{m}) \simeq 8\times 10^{-23}$, and compare it to the optimal sensitivities of several \ac{GW} detectors.
These anisotropies can be observed at low multipoles by \ac{LISA} and \ac{ET}-\ac{CE} network, and at up to much larger multipoles by \ac{BBO} and \ac{DECIGO}. 

The anisotropies in the \ac{GWB} reveal the $CP$-violating dynamics in \ac{AD} baryogenesis.
In Eq.~(\ref{eq:Cl}), given $\cA_\mathrm{L}\propto \cot^2{n\theta_\mathrm{i}}$, $F_\mathrm{nl}\propto \tan^2{n\theta_\mathrm{i}}$, and $\epsilon\sim 1$, the reduced angular power spectrum exhibits a basic relation to the $CP$ phase of $\phi$, i.e., $\ell(\ell+1)\, \widetilde{C}_{\ell}\propto \tan^2{n\theta_\mathrm{i}}$.
The measurement of the \ac{GW} anisotropies, along with the baryonic isocurvature perturbations on \ac{CMB} scales, provides two independent probes to the parameter space of $\{\theta_\mathrm{i}, H_\mathrm{i}/|\phi_\mathrm{i}| \}$.
For instance, the results $\ell(\ell+1)\, \widetilde{C}_{\ell}\, \simeq 0.016$ and $\cA_\mathrm{L}\simeq 10^{-11}$ given in the last paragraph indicate $\tan{n\theta_\mathrm{i}}\simeq 70.7 $ and $|\phi_\mathrm{i}|\simeq 4.3\times 10^3 H_\mathrm{i}$ for $n=6$ in our model.
This $CP$ violation also closely relates the observed baryon asymmetry through $n_b\propto \sin{n\theta_\mathrm{i}}$.
The large anisotropies imply sizable $CP$ violation, which is crucial for successful baryogenesis.
We expect that future \ac{GW} observations could detect these essential parameters for \ac{AD} baryogenesis, and provide profound implications for the origin of baryon asymmetry.

\emph{Conclusion and discussion.}---In this work, we studied the anisotropies in the \ac{GWB} related to \ac{AD} baryogenesis as a novel observable for this high-energy mechanism.
The non-Gaussianity in baryonic isocurvature perturbations from \ac{AD} baryogenesis couples large- and small-scale Q-ball density perturbations, thus leading to large-scale anisotropies in the \ac{GWB} sourced by Q-balls.
Resulting from large $|F_\mathrm{nl}|$ and large $\cN$ in this mechanism, the anisotropies in the \ac{GWB} can be significant with $\ell(\ell+1)\, \widetilde{C}_{\ell}\sim \cO(10^{-2})$, and are potential to detected by future \ac{GW} experiments like \ac{LISA}.
These anisotropies serve as a powerful probe to detect \ac{AD} baryogenesis and the $CP$-violating dynamics of \ac{AD} field, with valuable physical insights into the problem of baryon asymmetry.

Compared with other sources, these \ac{GW} anisotropies provide distinctive signals for \ac{AD} baryogenesis.
(a) Multipole dependence: The angular power spectrum scales as ${C}_\ell \sim \ell^{-2}$ in our model, contrasting with the ${C}_\ell \sim \ell^{-1}$ scaling for astrophysical sources (see Ref.~\cite{LISACosmologyWorkingGroup:2022kbp} for a review).
(b) Cross-correlation with the \ac{CMB}: These anisotropies showcase minimal cross-correlation with the \ac{CMB} due to the isocurvature nature of the \ac{GW} sources, distinguishing them from anisotropies produced by adiabatic sources \cite{Luo:2025lgr}.
(c) Large anisotropies: The observation of the anisotropies  $\ell(\ell+1)\,\widetilde{C}_\ell\sim 10^{-2}$ requires not only highly non-Gaussian sources but also a rapid \ac{eMD}-to-\ac{RD} transition (for comparison, see the results from scenarios without such transition in Refs.~\cite{Bartolo:2019zvb,Li:2023qua,Li:2023xtl,Schulze:2023ich,Wang:2023ost,Yu:2023jrs,Ruiz:2024weh}).
These features narrow down the \ac{GW} origins to limited scenarios, particularly Q-balls in \ac{AD} baryogenesis, offering crucial clues to validate this mechanism.

We demonstrated that large anisotropies can be generated in \ac{AD} baryogenesis even in the linear regime. 
If the \ac{eMD} era lasts sufficiently long, $\delta_\mathrm{Q}$ would exceed the unit and lead to more violent \ac{GW} production, suggested by numerical studies \cite{Kawasaki:2023rfx,Dalianis:2020gup,Eggemeier:2022gyo,Fernandez:2023ddy,Dalianis:2024kjr}.
The study of the \ac{GW} anisotropies beyond the linear regime may provide insights into \ac{AD} baryogenesis in more general cases, and we leave it for future investigation.

Recent studies indicate that some details in Q-ball models (e.g., the speed of decay \cite{Inomata:2019ivs,Inomata:2019zqy,Pearce:2023kxp,Pearce:2025ywc,Zeng:2025ecx} and the mass distribution \cite{Inomata:2020lmk,Pearce:2025ywc}) can significantly affect $\bar{\Omega}_\mathrm{gw}$.
In contrast, we emphasize that $\widetilde{C}_\ell$ is insensitive to these details since they meanwhile affect the numerator and denominator of $\epsilon$ in Eq.~(\ref{eq:Cl}).
This insensitivity can be understood as $\widetilde{C}_\ell$ only cares about the physics related to large scales, which usually originates from the initial conditions of $\phi$ during inflation.
Therefore, the relation between the $CP$ violation in \ac{AD} baryogenesis and the \ac{GW} anisotropies remains robust.

Our work established a framework to probe the $CP$ violation through \ac{GW} anisotropies.
Future works could include more scenarios, e.g., Q-balls with non-linear mass-charge relation \cite{Kusenko:1997si,Dvali:1997qv,Enqvist:2003gh}, the presence of anti-Q-balls \cite{Hiramatsu:2010dx}, or Q-ball modulated reheating \cite{Kamada:2010yz}, etc. 
These effects may generate additional perturbations and non-Gaussianity, making the comprehensive analysis of $\cS$ more complicated.
However, for the calculation of \ac{GW} anisotropies, we only need to modify the specific expression of $F_\mathrm{nl}$, without altering the framework of this paper.
Besides, this work can be extended to other baryogenesis mechanisms with non-Gaussian baryonic perturbations and Q-balls, e.g., spontaneous baryogenesis \cite{Cohen:1988kt,Chiba:2003vp,Takahashi:2003db}.

In light of the recent observation of the $CP$ violation in baryon decays at the LHCb \cite{LHCb:2025ray}, we are entering a new era for searching additional $CP$ violation sources beyond the \ac{SM} and determining the nature of baryon asymmetry.
In the future, \ac{GW} anisotropies are expected to detect the $CP$ violation at energy scales far beyond the reach of traditional particle physics experiments, opening a new road to address these essential problems.  
\\

We appreciate Jun-Peng Li for helpful discussion. 
This work is supported by the National Key R\&D Program of China No. 2023YFC2206403 and the National Natural Science Foundation of China (Grant No. 12175243).
\\
\\



\bibliography{AD}

\end{document}